# Low Barrier Nanomagnet Design for Binary Stochastic Neurons: Design Challenges for Real Nanomagnets with Fabrication Defects

Md Ahsanul Abeed[1,*], and Supriyo Bandyopadhyay[1,**]

[1] Department of Electrical and Computer Engineering, Virginia Commonwealth University, Richmond, VA 23284, USA
* Student Member, IEEE
** Fellow, IEEE

Abstract—Much attention has been focused on the design of low barrier nanomagnets (LBM), whose magnetizations vary randomly in time owing to thermal noise, for use in binary stochastic neurons (BSN) which are hardware accelerators for machine learning. The performance of BSNs depend on two important parameters: the correlation time $\tau_c$ associated with the random magnetization dynamics in a LBM, and the spin-polarized pinning current $I_p$ which stabilizes the magnetization of a LBM in a chosen direction within a chosen time. Here, we show that common fabrication defects in LBMs make these two parameters unpredictable since they are strongly sensitive to the defects. That makes the design of BSNs with real LBMs very challenging. Unless the LBMs are fabricated with extremely tight control, the BSNs which use them could be unreliable or suffer from poor yield.

## I. INTRODUCTION

Binary nanomagnets, with two stable magnetization directions, have long been thought of as potential replacements for transistor switches because they can be more energy-efficient in some circumstances and have the added boon of being non-volatile. They are switched between their two stable states with either current-controlled mechanisms (e.g. spin transfer torque [1], domain wall motion [2], giant spin Hall effect [3], spin diffusion from topological insulators [4], spin-orbit torque [5], etc.) or voltage-controlled mechanisms (e.g. voltage-controlled-magnetic-anisotropy (VCMA) [6], straintronics [7], etc.). The latter are more energy-efficient, but also more error-prone. Theoretical estimates of switching errors in magneto-elastic switching (straintronics) range from $10^{-3}$ [8] to as low as $10^{-8}$ [9], but experiments show a much larger error probability exceeding 0.1 [10, 11]. Recently, we showed that this large difference between theory and experiments accrues from the fact that real nanomagnets used in experiments invariably have structural defects and they exacerbate the switching error [12]. The error probability associated with VCMA switching is also relatively large and experiments show it to be ~$10^{-5}$ [13] or more.

Error probabilities this high will probably make these switches unusable for Boolean logic, and maybe even memory if we insist on low write energies (lower write energies lead to higher error rates). In logic, errors propagate and consequently the switching error probability of a switch should be less than $10^{-15}$ for logic applications [14]. Memory is more forgiving since errors do not propagate, but in order to avoid repetitive write/read/re-write operations, the error probability still needs to be below $10^{-9}$ [15]. Such low error probabilities may be out of reach for voltage-controlled binary nanomagnetic switches, and maybe a tall order for even current-controlled (more dissipative) ones if we wish to expend no more than ~1 fJ of write energy per bit. Using higher write energies might improve the write error probability, but high energies are undesirable in embedded applications.

Because of this impasse, there has recently been increasing interest in applying magnetic switches for *non-Boolean* applications such as in Bayesian inference engines [16, 17], restricted Boltzmann machines [18], image processing [19, 20], computer vision [21], neural networks [22, 23], analog-to-digital converters [24], and probabilistic computing [25, 26]. These applications can tolerate much larger error probabilities than Boolean logic or memory can. Probabilistic computing has attracted particular attention since it can leverage the thermally induced fluctuations in the magnetization of a nanomagnet (which would have caused unwanted errors in logic or memory devices) to actually elicit useful computational activity. Stochastic binary neurons are a class of probabilistic computing devices. They are implemented with low energy barrier nanomagnets (LBMs) where the energy barrier separating the two stable magnetization states is purposely made small enough that thermal noise can make the magnetization fluctuate randomly with time. This random magnetization distribution is utilized for computation [24, 27, 28]. One way to realize LBMs is to fashion them out of nearly circular ultrathin disks of small cross-section, resulting in low shape anisotropy energy barrier on the order of the thermal energy kT. These nanomagnets have in-plane magnetic anisotropy (IMA) and they may have an advantage over those with perpendicular magnetic anisotropy in BSNs [29]. If we monitor the normalized magnetization component $m_z(t)$ along any one direction (say, the z-direction) on the surface of the LBM, it will randomly vary between -1 and +1 with time $t$ because of thermal noise.

Ref. [29] recently discussed the design of LBMs for binary stochastic neurons (BSN) and focused on two important parameters that govern the BSN operation: the correlation time $\tau_c$ and the pinning current $I_p$. The former is the *full width at half maximum* of the temporal autocorrelation function which we define as

$$C(\tau) = \int_{-\infty}^{\infty} dt \left[ m_z(t) m_z(t+\tau) \right], \quad (1)$$

where $m_z(t)$ is the magnetization component along any chosen direction (say, the z-direction) at any time $t$. The latter is the magnitude of the spin-polarized current needed to pin the magnetization in the direction of the spin-polarization of the current due to spin transfer torque within a specified time. These two parameters are crucial in designing LBMs for BSNs.

The purpose of this paper is to explore how sensitive $\tau_c$ and $I_p$ are to small *fabrication defects* in real LBMs. Strong sensitivity will make the design of LBMs for BSNs extremely challenging since $\tau_c$ and $I_p$ would become unpredictable, given that defects are unpredictable. Particularly, the response time of BSNs, which is dependent on $\tau_c$, will vary randomly if the fabrication process is not well-controlled. This could be debilitating for BSN applications of LBMs and make LBMs unsuitable for BSNs. This study is carried out to examine this possibility.

## II. SIMULATION OF LOW BARRIER NANOMAGNETS (LBM)

We have simulated the magnetization dynamics in a low barrier nanomagnet (LBM) made of cobalt. The LBM is a thin elliptical disk of small eccentricity (nearly circular) whose major axis dimension is 100 nm, minor axis dimension is 99.7 nm and thickness 6 nm. The calculated shape anisotropy energy barrier is 1.3 kT at room temperature because of the small eccentricity.

We start with the magnetization having components $[m_z(t=0) = 0.99; m_x(t=0) = 0.141; m_y(t=0) = 0]$ initially, with the $z$-axis being along the major axis, and simulate its temporal evolution in the presence of a random thermal field given by

$$\vec{H}_{th}(t) = H_x(t)\hat{x} + H_y(t)\hat{y} + H_z(t)\hat{z}$$
$$H_i(t) = \sqrt{\frac{2\alpha kT}{\gamma(1+\alpha^2)\mu_0 M_s \Omega \Delta t}} G_{0,1}^i(t) \quad (i = x, y, z), \quad (2)$$

where $\alpha$ is the Gilbert damping factor of cobalt ($\alpha = 0.01$), $G_{0,1}^i(t)$ is a Gaussian of zero mean and unit standard deviation, $\gamma = 2\mu_B \mu_0 / \hbar$, $\mu_B$ is the Bohr magneton, $\mu_0$ is the permeability of free space, $M_s$ is the saturation magnetization of cobalt ($1.1 \times 10^6$ A/m) [30], $\Omega$ is the nanomagnet volume and $\Delta t$ is the time step used in the simulation (0.1 ps). We ignore magneto-crystalline anisotropy, assuming that the nanomagnet is amorphous. The simulation is carried out with the micromagnetic simulator MuMax3 [31]. We assumed that the exchange constant of cobalt is $3 \times 10^{-11}$ J/m [32]. The simulation is carried out for 1 ns, which provides more than enough statistics to calculate the auto-correlation function in Equation (1). All calculations are carried out for room temperature (300 K).

In our study, we simulate the temporal dynamics of the magnetization in a defect-free nanomagnet (C0) and nanomagnets with six different types of defects (C1-C6) shown in Fig. 1. These defects are commonplace in nanomagnets that have been fabricated in our lab with electron-beam patterning of a resist, followed by development, evaporation of (ferromagnetic) metal and lift-off. Fig. 2 shows atomic force micrographs of such fabricated nanomagnets. The idealizations in Fig. 1 closely approximate some of the defects observed experimentally and shown in Fig. 2.

The MuMax3 simulations allow us to determine the normalized magnetization component $m_z(t)$ at any instant of time $t$ in any nanomagnet, and that allows us to calculate all quantities of interest.

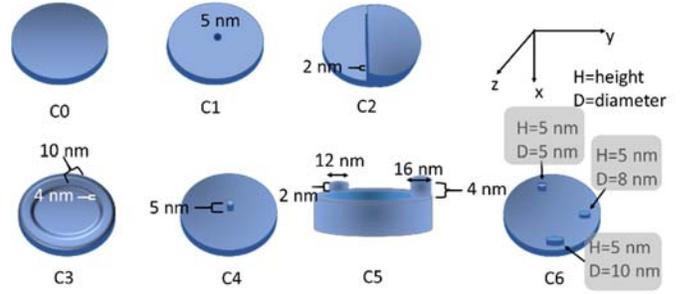

Fig. 1: Defect-free and defective nanomagnets. The dimensions of the nanomagnets are: major axis = 100 nm, minor axis = 99.7 nm, thickness = 6 nm. C1 has a 5-nm diameter hole in the center. C2 has two halves with two different thicknesses – one with a thickness of 6 nm and the other with a thickness of 8 nm. C3 has an annulus along the circumference of width 10 nm and height 4 nm. C4 has a cylindrical hillock in the center of diameter 5 nm and height 5 nm. C5 is similar to C3, except the width and height of the annulus varies at two arbitrary locations. In one location the annulus is replaced by a cylinder 8 nm tall and 16 nm in diameter and in another location by a cylinder 6 nm tall and 12 nm in diameter. This more closely approximates the features observed under an atomic force microscope and shown in the right panel of Fig. 2. The defect C6 consists of cylinders of arbitrary dimensions located at arbitrary locations on the surface. This emulates random surface roughness. This figure is not to scale.

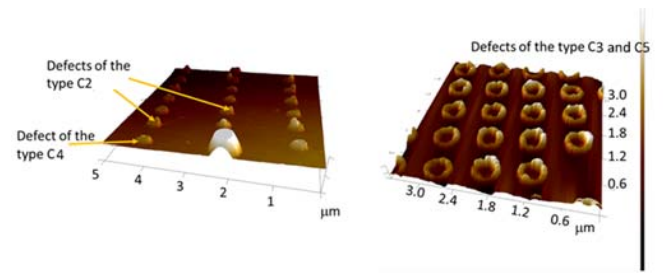

Fig. 2: Atomic force micrographs of defects observed in different fabrication runs in our lab. The simulated defects in Fig. 1 approximate these observed defects. Reproduced from [12] with permission.

## III.  RESULTS

In Fig. 3, we plot the autocorrelation function [Equation (1)] for a defect-free nanomagnet (C0) along with those for nanomagnets with defects C1-C6. Clearly, there are vast differences between the auto-correlation functions for a defect-free nanomagnet and some defective nanomagnets. Table I lists the correlation times $\tau_c$ calculated for the defect-free and defective nanomagnets.

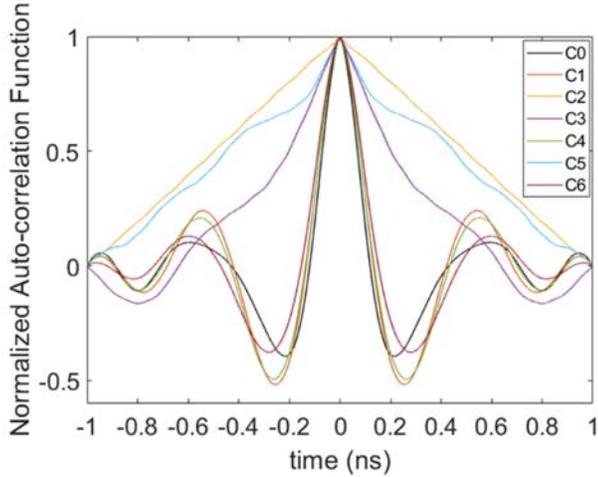

Fig. 3: The normalized temporal auto-correlation functions (defined in Equation (1)) plotted as a function of time for defect-free and defective LBMs. All LBMs in this plot are nominally identical, except they have different defects. Their auto-correlation functions are very different – both qualitatively and quantitatively – showing that this parameter is unpredictable in real LBMs that have random defects. Single (C1 and C4), and distributed (C6) *localized* defects do not change the auto-correlation functions by much, but *delocalized* (extended) defects (C2, C3 and C5) affect the auto-correlation functions drastically.

Table I: Correlation times $\tau_c$ for various defects

| Type of defect | Correlation time (ns) |
|---|---|
| C0 | 0.139 |
| C1 | 0.158 |
| C2 | 0.995 |
| C3 | 0.403 |
| C4 | 0.157 |
| C5 | 0.905 |
| C6 | 0.173 |

We have also found the magnitudes of the (perpendicular-to-plane) spin-polarized current which stabilizes (or pins) the magnetization in the direction of the spin polarization of the current owing to spin transfer torque. The pinning current will depend on the time allocated for the pinning to occur, which we have arbitrarily chosen to be 5 ns. We have found the pinning currents for the defect-free and defective nanomagnets to assess how much defects influence the pinning current.

To find the pinning currents, we pass a 100% spin polarized current, whose spin polarization is in the –z direction, perpendicular to the plane of the nanomagnet. The current is assumed to retain its spin polarization throughout the magnet since the spin diffusion length in cobalt at room temperature is $38 \pm 12$ nm [33] and the magnet thickness is only 6 nm. We pass the current for 5 ns (which would be the typical clock period in a neural circuit) and plot the z-component of magnetization as a function of the current magnitude at the completion of 5 ns of current passage (i.e. at the end of the assumed clock period). This is shown in Figure 4. The pinning current is the minimum current that can pin the magnetization close to the –z-direction (direction of the current's spin polarization) within the allotted 5 ns. At low values of the current, the magnetization component does not settle to the –z-direction definitively after 5 ns, but it does so beyond a threshold current (except for some defects), and this threshold current is the "pinning current".

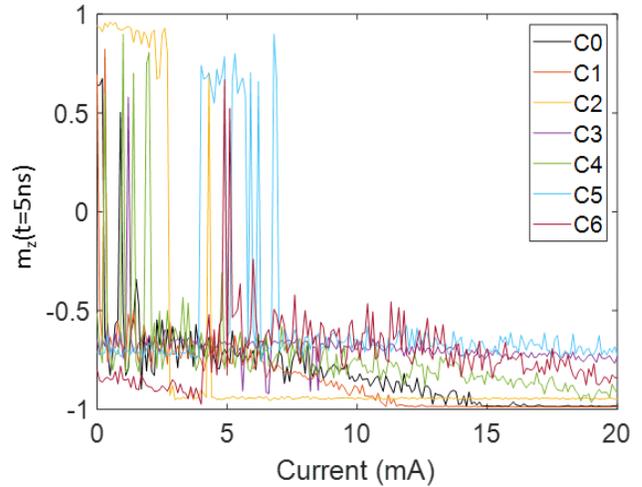

Fig. 4: Normalized magnetization component in the direction opposite to that of current spin polarization after 5 ns of current passage as a function of the current magnitude for defect-free and defective nanomagnets.

In Fig. 4, we see that the normalized magnetization component in the direction of pinning at the end of 5 ns of current passage [$m_z(t=5$ ns)] fluctuates (between almost -1 and +1) at low current levels, i.e. the magnetization does not pin *monotonically* with increasing current. This feature is not just due to thermal noise, although noise also plays a role. We have verified that similar non-monotonicity (although much more muted) is present even at 0 K temperature when no thermal noise is present. It is a consequence of the magneto-dynamics which consists of precession and damping. When the current is turned on, the magnetization begins to precess and gradually align along the pinning direction owing to the damping caused by the current. At low current levels, the damping is weak and hence the precession would not have died out within 5 ns. If we take a snapshot at the end of 5 ns, the magnetization component in the pinning direction [$m_z(t=5$ ns)] will depend on what angle the magnetization has rotated through at the end of 5 ns and that depends on the interplay between the precessional and damping dynamics, which does not have a monotonic dependence on current. As a result, $m_z(t=5$ ns) need not vary monotonically with current and the pinning need not occur monotonically. Thermal noise exacerbates this effect, resulting in the low current fluctuations seen in Fig. 4. At high current levels (large Slonczewski torque), the damping is strong enough to have aligned the magnetization in the pinning direction after 5 ns, which is why we would not see the fluctuations at higher current.

In Fig. 4, we observe that for defects C3 and C5, the magnetization does not settle in the direction of the current's spin polarization, i.e. $m_z \approx -1$, after 5 ns, even when the current is as large as 20 mA. Instead the z-component of the magnetization seems to saturate to a value that is approximately – 0.7, after the current exceeds ~8 mA. The y-component however does not saturate and continues to change if we vary the current (no pinning). For defect C6, the z-component does not settle into any value up to a current of 20 mA, but clearly trends towards a value of -1 as the current is increased (it might ultimately settle at close to -1, i.e. get pinned, if we exceed a current of 20 mA). For all other defects, (and the defect-free nanomagnet), the z-component of the magnetization does settle to a value close to -1 (indicating that the magnetization has been pinned in the direction of the current's spin polarization) after the current exceeds a threshold value.

Clearly, for defects C3 and C5 (C5 is similar to C3 in geometry), current up to 20 mA cannot pin the magnetization in the direction of the spin polarization of the current (i.e. make $m_z \approx -1$) in 5 ns, indicating that there may be no reasonable pinning current for these types of defect, if no more than 5 ns is allowed for pinning. These types of defects may not allow pinning within a reasonable time with a reasonable current. On the other hand, defect C6, which represents surface roughness, may allow pinning at the end of 5 ns, but would require a very large current (> 20 mA). Thus, the pinning current magnitude – or if pinning is even possible – within a given time depends on the nature of the defects. Defects can either increase or decrease the pinning current, which can vary over almost an order of magnitude because of the defects. This attests to the unpredictability of the pinning current when defects are present.

Table II lists the pinning currents (for 5 ns pinning duration) for various types of defects. They are determined from Fig. 4.

Table II: Pinning currents for various defects that pin within 5 ns

| Type of defect | Pinning current (mA) |
|---|---|
| C0 | 14.8 |
| C1 | 11.9 |
| C2 | 3.1 |
| C3 | … |
| C4 | 19 |
| C5 | … |
| C6 | >20 |

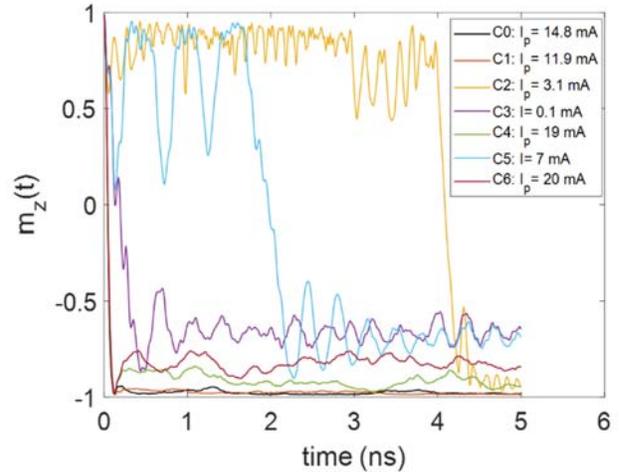

Fig. 5: Normalized magnetization component in the direction of spin polarization of the injected current as a function of time at different current magnitudes for the seven different types of nanomagnets (defective and defect-free).

In Fig. 5, we plot the time variation of the z-component of the magnetization (between 0 and 5 ns) for the defect-free and defective nanomagnets when the corresponding pinning current (taken from Table II) is injected to align the magnetization in the direction of the current's spin polarization. In the case of defects C3 and C5, there is no pinning. So, we use the minimum current that stabilizes the z-component of the magnetization (albeit not in the desired direction). Note that for all defects except C2 and C5, the z-component of the magnetization saturates to a steady-state value very quickly (<0.5 ns) when the pinning current is passed. The defect C2 takes much longer (>4.3 ns) and C5 takes > 2 ns. C2 has two different thicknesses in two different halves and is an extended (delocalized) defect like C3, C5 and distributed localized defects like C6. This plot shows that the "pinning time" is also affected by defects.

## IV. CONCLUSION

This work shows that defects can increase or decrease both the correlation time $\tau_c$ and the pinning current $I_P$ and make them vary randomly by nearly an order of magnitude. That makes the parameters of BSNs designed with LBMs very difficult to control. Either LBMs will have to be fabricated with extremely tight control (defect-free) or they may suffer from poor yield. Ref. [29] discussed a different genre of BSNs that are controlled more by transistor characteristics than nanomagnet characteristics. They may be more immune to fabrication defects than those which rely critically on nanomagnet characteristics.

To conclude, this study has shown that unavoidable fabrication defects may introduce wide variability in the characteristics of LBMs that are utilized in BSNs [29] or other constructs like analog-to-digital converters [24]. While stochastic computing is generally more tolerant of variabilities than deterministic computing (e.g. Boolean computing) [34], whether variability spanning an order of magnitude is tolerable remains an open question and needs to be addressed. Some defects are more harmful than others; for example, the defect of type C3 (rim around the circumference) or C5 (which is similar to C3), does not even allow one to pin the magnetization with any reasonable current in a reasonable time. Generally speaking, extended (delocalized) defects like C2, C3, C5 affect the auto-correlation function much more than localized defects like C1, C4 and C6, and consequently they have a much more severe effect on the correlation time. Delocalized defects may be less common than the localized ones, but their effect is also much more debilitating, and that makes them more of a concern than localized defects.

## ACKNOWLEDGMENT

We acknowledge Prof. Supriyo Datta of Purdue University for fruitful discussions. This work was supported in part by the U.S. National Science Foundation under grants ECCS-1609303 and CCF-1815033.